# Älypuhelin ja kasvokkaisen vuorovaikutuksen muuttuvat käytänteet

Sanna Raudaskoski, Eerik Mantere ja Satu Valkonen

## Abstrakti

Älypuhelimen käyttö on lisääntynyt nopeasti. Se on henkilökohtainen esine ja kulkee yleensä aina mukana. Siksi sen käyttö voi sijoittua keskelle melkein mitä tahansa vuorovaikutuskontekstia, varsinkin kun laite usein myös kutsuu äänen, värinän tai visuaalisten efektien avulla omistajaansa. Jatkuvasti käsillä olevat älypuhelimet ovat vaikuttaneet myös kasvokkaisen keskustelun käytänteisiin. Tarkastelemme artikkelissamme, miten älypuhelinten käyttö muokkaa samaan aikaan tapahtuvaa kasvokkaista vuorovaikutusta ja millaisia uusia aspekteja se siihen tuo. Olemme luoneet kaksi uutta käsitettä kuvaamaan näitä aspekteja: 1) *tahmea medialaite* ja 2) *sivustakatsojan pimento*. Tahmea medialaite kuvastaa tilanteita, joissa älypuhelinta käyttävää henkilöä on hankala saada mukaan tai pitää mukana kasvokkaisessa keskustelussa. Hänen huomionsa ei helposti irtoa "tahmeasta" laitteesta, tai se palaa siihen nopeasti takaisin. Sivustakatsojan pimennolla tarkoitamme sitä, että sivustakatsojan on vaikea tietää, mitä älypuhelimen käyttäjä pieniruutuisella laitteellaan tekee tai sitä, missä vaiheessa tekeminen on ja onko hänen toimintaansa soveliasta keskeyttää. Tutkimuksemme perustuu etnometodologiaan ja etnometodologiseen vuorovaikutuksen analyysiin, mutta sovellamme tutkimuksissamme myös eläytymismenetelmää ja kvantitatiivisia tutkimusmenetelmiä.

Asiasanat: etnometodologia, kasvokkainen vuorovaikutus, sivustakatsojan pimento, tahmea medialaite, älypuhelin

# Smartphone and the changing practices of face-to-face interaction

## Abstract

Smartphone use has grown rapidly, but the ways it shapes concurrent face-to-face interaction remains scarcely studied. In our research we have formulated two new concepts to depict this: 1) *Sticky media device* illustrates situations in which a person using a screen media device is difficult to get fully involved with ongoing face-to-face conversation. Their attention is not easily removed from the "sticky" device or returns to it quickly even if it is momentarily removed. This article adds to the theoretical underpinnings of the concept that we previously described mainly empirically. By 2) *bystander inaccessibility* we mean the difficulty of a bystander to a smartphone user to be aware of what kind of action the user is undertaking with the device and what the phase of the activity is. Our research is based on the theory of ethnomethodology. In addition to ethnomethodological analysis of interaction, we also apply other reseach methods. We illustrate the phenomena of sticky media device and bystander inaccessibility by analyzing 1) naturalistic video data, 2) written role playing materials and 3) quantitative data, all of which concentrate on the overlapping of smartphone use and face-to-face conversation.

## Johdanto

Erving Goffmanin (1964) mukaan sosiaalisessa tilanteessa vähintään kaksi ihmistä ovat vastavuoroisesti tietoisia toistensa läsnäolosta. Hänen mukaansa vuorovaikutusta ei kuitenkaan voida tyydyttävästi ymmärtää ilman sosiaalisen tilanteen fyysisten aspektien huomioon ottamista. Myös Harold Garfinkel (1967) korosti vuorovaikutuksen ympäristön merkitystä ja painotti kaiken sosiaalisen toiminnan saavan merkityksensä kontekstinsa kautta. Garfinkel näki vuorovaikutuksen tutkimuksen koko yhteiskunnan kannalta olennaisena tehtävänä, sillä sosiaaliselle toiminnalle välttämätön jaettu ymmärrys rakennetaan nimenomaan arkisessa vuorovaikutuksessa (Garfinkel 2002; Heritage 1984).

Garfinkelin luomassa etnometodologisessa tutkimusperinteessä sosiaalista toimintaa selitetään tarkastelemalla niitä yleensä itsestäänselvyyksinä pidettyjä seikkoja, joita hyödyntämällä ihmiset tuottavat ja tunnistavat sosiaalisia toimintoja. Esimerkiksi keskustelu noudattaa tarkkoja sääntöjä, joiden kautta ihmiset pitävät keskustelua yllä, ymmärtävät toisiaan sekä tulevat ymmärretyksi (Garfinkel 1963; 1967). Etnometodologian pohjalle syntyneen keskustelunanalyysin mukaan keskustelun dynamiikka pohjautuu erilaisiin yhteisiin toimintanormeihin, joihin on sisäänkirjoitettuna myös vuorovaikutuksen moraalinen koodisto.

Tässä artikkelissa tutkimme niitä aspekteja, joita kaikkialla läsnä olevat älypuhelimet ovat tuoneet kasvokkaiseen vuorovaikutukseen. Seuraten Goffmanin ja Garfinkelin näkemystä yhteiskuntatieteellisesti relevantista tutkimuksesta, tuomme vuorovaikutuksen tutkimukseen uusia käsitteitä, joiden avulla sosiaalista toimintaa voidaan tehdä tässä ajassa ymmärrettäväksi.

Mediavälitteistä vuorovaikutusta on tutkittu paljon (esim. Baym 2015; Castells ym. 2007; Foth ym. 2011; Ling & Campell 2008; Katz 2008). Vaikka myös oma tutkimuksemme tietyllä tapaa käsittelee vuorovaikutteisen median käyttöä, tutkimuksemme varsinaisena kohteena on kuitenkin kasvokkainen vuorovaikutus. Lähtökohtanamme ei ole mediavälitteisen ja kasvokkaisen vuorovaikutuksen vastakkainasettelu ja vertailu keskenään (vrt. Arminen, Licoppe & Spagnolli 2016). Keskitymme siihen, miten ruutumedian, erityisesti älypuhelimen, käyttö vaikuttaa kasvokkaiseen vuorovaikutukseen niiden henkilöiden näkökulmasta, jotka jakavat laitteen käyttäjän kanssa saman fyysisen tilan. Tutkimuskohteenamme on siis sosiaalinen tilanne sen klassisessa goffmanilaisessa merkityksessä (Goffman 1964). Tämän tyyppiselle tutkimukselle on tarvetta, sillä vastaavanlaista teoreettista pohdintaa, empiiristä tutkimusta tai ilmiöiden käsitteellistämistä ei ole vielä muualla tehty aikakautena, jona älypuhelinten käyttö on mukana arkielämän mitä erilaisimmissa tilanteissa.

Käsitteellistämme älypuhelinajan vuorovaikutuksen uusia aspekteja kahden termin avulla: 1) *tahmea medialaite* ja 2) *sivustakatsojan pimento*. Tahmean medialaitteen käsitteellä kuvaamme tilannetta, jossa jotain ruutumedialaitetta, usein älypuhelinta, käyttävää ihmistä on hankala saada mukaan kasvokkaiseen vuorovaikutukseen. Hänen huomionsa ei helposti irtoa "tahmeasta" laitteesta. Sivustakatsojan pimennon käsitteellä tarkoitamme sitä, että sivustakatsojan on vaikea tietää, millaisia toimintoja etenkin älypuhelimen käyttäjä kohtalaisen pieniruutuisella laitteellaan on tekemässä. Olemme tutkineet näitä ilmiöitä erityisesti älypuhelimen käyttötilanteissa, mutta käsitteet ovat sovellettavissa yleisestikin tilanteisiin, joissa ruutumedialaitteita käytetään. Tämän artikkelin tutkimustehtävänä on luoda synteesiä aiempien tutkimustemme (Mantere & Raudaskoski 2015; 2017; Raudaskoski, Mantere & Valkonen 2017; Mantere, Raudaskoski & Valkonen 2018) tuloksista mutta myös esittää uutta empiiristä analyysiä aiempien teoreettisten mallinnustemme tueksi.

Olemme selittäneet tahmean medialaitteen vuorovaikutuksellista dynamiikkaa keskustelunanalyysin avulla muissa tutkimusjulkaisuissa (Mantere & Raudaskoski 2015; 2017), joten emme tässä yhteydessä enää palaa noihin tarkkoihin vuorovaikutusanalyyseihimme. Esittelemme käsitteen tässä artikkelissa yleisemmällä tasolla, sillä sen ymmärtäminen on keskeistä myös hahmotettaessa sivustakatsojan pimennon käsitettä. Sivustakatsojan pimennon käsitettä olemme puolestaan käsitelleet aiemmin vain teoreettisesti (Raudaskoski ym. 2017). Tässä artikkelissa raportoimme sivustakatsojan pimentoa testaavaa uutta empiiristä tutkimustamme, jossa hyödynsimme etnometodologiaa, eläytymismenetelmää sekä perinteistä kyselytutkimusta. Artikkelissa käsittelemme myös nuorten ja nuorten aikuisten videotallennettuja keskusteluja, joita analysoimme soveltaen keskustelunanalyysiä. Lisäksi analysoimme uudentyyppisellä eläytymismenetelmällä, videoaineistoihin eläytymisellä, kerättyjä kirjoituksia, jotka käsittelevät nuorten älypuhelinten käyttöä kasvokkaisissa vuorovaikutustilanteissa.

Ennen keskeisiin käsitteisiin paneutumista tarkastelemme matkapuhelinten teknologioiden muutokseen, matkapuhelinten yleistymiseen sekä henkilökohtaiseen saavutettavuuteen liittyviä tekijöitä. Ne ovat yhteiskunnassa laajemmin tapahtuneita muutoksia, jotka ovat osaltaan mahdollistaneet älypuhelinajan uudenlaisia vuorovaikutuskäytänteitä.

## Matkapuhelimiin liittyvät uudet vuorovaikutuksen konventiot

Digitaalisen viestinnän kehittyminen on yksi lähihistorian merkittävimpiä muutoksia jokapäiväisen vuorovaikutuksen käytänteissä. Matkapuhelinten sekä viime vuosina datayhteyksillä varustettujen älypuhelinten käyttöönotto on tämän muutoksen silmiinpistävimpiä edustajia. Jo 1990-luvulla soittamisen ja tekstiviestit mahdollistaneet kännykät vaikuttivat radikaalisti arjen käytänteisiin (esim. Ling 2004). Samaan aikaan ihmiset oppivat käyttämään tietokoneillaan internetin palveluita, kuten sähköpostia, ja hieman myöhemmin myös sosiaalisen median sovelluksia. Kun 2000-luvun loppupuolella uuden sukupolven matkapuhelinteknologia yhdisti pieneen kannettavaan laitteeseen tehokkaat datayhteydet, matkapuhelinten kautta ryhdyttiin käyttämään yhä enemmän internetin sovelluksia.

Matkapuhelimet ovat enenevässä määrin kaikkialla maailmassa henkilökohtaisesti käytettyjä välineitä, joista on tullut osa jokaisen ihmisen arkipäivää (Castells ym. 2007; Kakihara 2014; Katz 2008). Vielä 2000-luvun alussa mobiiliteknologian kehittäjien visiot kännyköiden kehityksestä olivat keskenään hyvin erilaisia. Osa ennusti, että kännykästä häviää ruutu ja siitä tulee vain pieni akullinen laite, joka hoitaa tarvittavat puhelinyhteydet erillisille mikrofonilla varustetuille kuulokkeille sekä datayhteydet erilliselle kannettavalle tietokoneelle. Epäiltiin, etteivät ihmiset halua kantaa mukanaan yhtä isohkoa laitetta, johon olisi integroituneena kaikki silloin näköpiirissä olleet mahdolliset käyttötavat. (Raudaskoski & Arminen 2003, 31–33.) Puhelimen ruutu ei kuitenkaan hävinnyt, vaan siitä on tullut uusien älypuhelimien keskeinen tekninen ominaisuus.

Kosketusnäytön ansiosta erillisiä toimintonäppäimiä ei enää tarvita ja suurin osa koko laitteen pinta-alasta koostuu ruudusta. Tapauskohtaisesti mukautuvat virtuaaliset painikkeet mahdollistavat lukemattoman määrän erilaisia käyttötarkoituksia, ja älypuhelimien käyttö sekä siihen saatavilla olevien sovellusten määrä ovatkin lisääntyneet voimakkaasti 2010-luvulla. Nykyään alle 54-vuotiaista 90 prosentilla ja 65–74-vuotiaistakin yli puolella (59 %) on älypuhelin. (Tilastokeskus 2018.) Lapsista lähes kaikilla 9-vuotiailla ja tätä vanhemmilla on käytössään älypuhelin (DNA 2017). Älypuhelin on suosituin laite internetin käyttöön, ja etenkin nuoret käyttävät internetiä pääasiassa ainoastaan älypuhelimen kautta (DNA 2017; eBrand 2016; Tilastokeskus 2017; 2018). Tämän artikkelin tutkimuskysymyksen kannalta on merkittävää, että koska laitteiden käyttö ja laitevälitteinen vuorovaikutus ovat voimakkaasti lisääntyneet, kaikkialla mukana olevat älypuhelimet ovat usein läsnä myös silloin, kun käymme kasvokkaisia keskusteluja.

Aina saavutettavissa oleminen on nykypäivän normaalia. Esimerkiksi nuoret ajattelevat, että ystäviin voi olla yhteydessä myös myöhään yöllä. Samoin älypuhelimen käyttöä vaikkapa ruokailun yhteydessä ei enää pidetä erityisen sopimattomana. (Kauppinen, Kivikoski & Manninen 2014, 32–34.) Myös työasioita hoidetaan niin myöhäisinä arki-iltoina kuin viikonloppunakin: yhä useamman ihmisen työtehtävät vaativat erilaisiin viesteihin reagoimista virka-ajan ulkopuolellakin (esim. Gant & Kiesler 2002; Wajcman, Bittman & Brown 2009). Suurin osa alle 12-vuotiaiden lasten vanhemmista hyväksyy älypuhelimen käytön henkilökohtaisten tai työhön liittyvien sähköpostien, pikaviestien tai tekstiviestien lähettämiseen, uutisten tai sosiaalisen median ilmoitusten seuraamiseen taikka laskujen maksamiseen samaan aikaan kun aikaa vietetään lasten kanssa (Kauppinen, Kivikoski & Manninen 2014, 31, 33). Jotkut vanhemmat kertovat, että älypuhelimen kautta he kuitenkin saattavat tempautua takaisin työmaailmaan, eikä siirtyminen perheen kanssa olemiseen aina tapahdu helposti (Radesky ym. 2016). Yhdysvaltalaisessa korkeakouluopiskelijoiden keskuudessa tehdyssä tutkimuksessa puolestaan kävi ilmi, että lähes 90 prosenttia opiskelijoista nukkuu puhelin käden etäisyydellä ja tuntee hereillä olonsa turvallisemmaksi, jos puhelin on heillä kädessä (Emanuel ym. 2015, 294–295).

Oletus jatkuvasta saatavilla olemisesta ei ole syntynyt vasta älypuhelimen aikakaudella. Jo puhelut ja tekstiviestit mahdollistaneet aiempien sukupolvien matkapuhelimet loivat kulttuurin, jossa toisten oletettiin reagoivan viesteihin ja soittoihin nopeasti. Vuonna 2005 Suomessa kerättiin lähinnä nuorten aikuisten lähettämistä tekstiviesteistä koostuva tekstiviestiaineisto, joka osoitti, että vastaukset viesteihin tulivat keskimäärin alle seitsemässä minuutissa. Lisäksi jos viestit viipyivät, ne alkoivat usein selonteolla siitä, mistä viipyminen johtui. (Raudaskoski 2009, 122–123.) Tanskalaisessa tekstiviestitutkimuksessa huomattiin, että jos vastaamiseen kului yli seitsemän minuuttia, 14-vuotiaat nuoret lähettivät uuden viestin muistutukseksi (Laursen 2005, 57). Jo tekstiviestiaikana jatkuvaa saatavilla oloa pidettiin niin itsestään selvänä, että kylpyyn menemisestä saatettiin ilmoittaa erikseen viestillä (Ito & Okabe 2005, 139). Nyt taas kosteutta kestäviä älypuhelimia pidetään usein mukana myös kylpyhuoneessa ja vessareissuilla (Rampton 2017).

Nykypäivän älypuhelinten yhä moninaisemmat toiminnot ja yhteydenpidon mahdollisuudet ovat monella tapaa luonteva jatkumo jo aiemmin olemassa olleelle matkapuhelimien käyttökulttuurille ja siihen liittyneille yhteisöllisille odotuksille. Sherry Turklen (2008, 122) mukaan olemme nyky-yhteiskunnassa kuin kahlittuja mukana kulkeviin puhelimiimme. Tavoittamattomissa olemalla rikotaan yhteisöllisesti muotoutuneita pelisääntöjä eikä tarjota muiden käyttöön kaikkia niitä vuorovaikutuksellisia resursseja, joita meillä oletetaan olevan.

Enää ei myöskään ajatella niin, että yhteydenottajan tulisi säädellä yhteydenpidon aikaa ja paikkaa. Vastuu digitaalisen median viestien virran hallitsemisesta on siirtynyt vastaanottajalle: hänen pitää asettaa laitteensa äänettömälle silloin, kun ei halua tulla häirityksi, tai jättää vastaamatta, jos on hankalassa paikassa. Kuitenkin pääsääntöisesti valveillaoloaikaan lähetettyihin viesteihin odotetaan pikaista vastausta, ja mikäli tärkeisiin yhteisöpalvelujen päivityksiin ei reagoida, henkilön oletetaan olevan toiminnastaan selontekovelvollinen. (vrt. Raudaskoski 2009, 68–69, 210.) Lisäksi laitetta "näprätään" myös pelkän tavan vuoksi, eli muulloinkin kuin asioita hoidettaessa. On tavallista, että esimerkiksi itselle tärkeiden yhteisöpalvelujen tilannetta käydään tarkistamassa tiheään tahtiin. (Oulasvirta ym. 2012.) Ruutua kannetaan aina mukana, ja ruutu on katsomista varten.

# Tahmea medialaite

> Kaverit eivät enää puhu toisilleen vaan tuijottavat kännykkäänsä, joka pysyy kädessä kuin liimattuna. Jos yrittää puhua jollekin, saa vain ärsyyntyneen vastauksen: Odota nyt. (Kahdeksasluokkalaisen tytön yleisönosastokirjoituksesta Aamulehdessä 10.10.2016)

Älypuhelimen käyttö voi usein tapahtua keskellä kasvokkaista vuorovaikutusta. Keskustelun ja muun sosiaalisen toiminnan perustavanlaatuisia rakennuspalikoita ja eteenpäin vieviä voimia ovat vuoroittaiset toimintaparit. Nämä voivat olla kehollisia tai verbaalisia, kuten esimerkiksi ojentaminen–vastaanottaminen tai kysymys–vastaus. Etnometodologian perustalle rakentuvan keskustelunanalyysin parissa tällaisia vuorottain tapahtuvia toimintoja kutsutaan vieruspareiksi. Vierusparijäsennys tarkoittaa sitä, että jokainen vuorovaikutuksellinen teko sisältää odotuksen jonkinlaisesta tekoon vastaamisesta. Tämän vastauksen on sijoituttava oikeaan paikkaan, jotta se muodostaisi edellä tapahtuneen kanssa yhteisesti ymmärrettävän vuorovaikutustoiminnon. Vierusparirakenteen rikkoutumista pidetään sosiaalisissa tilanteissa yleensä outona: jos et vastaa vaikkapa tervehdykseen, sitä kummeksutaan (Garfinkel 1963; Sacks, Schegloff & Jefferson 1974). Toiminnot saavat siten varsinaisen merkityksensä vuorovaikutuskontekstissa ja siinä toiminnan sekvenssissä, missä ne tapahtuvat. Keskinäisessä toiminnassa luotetaan pitkälti tähän järjestykseen ja säännönmukaisuuteen, joka yhteisön jäsenille on kokemuksen kautta tullut tutuksi. Esimerkiksi lapset omaksuvat sosiaalisia taitoja oppimalla juuri oikeanlaisten ja oikeaan paikkaan sijoittuvien vuorovaikutustekojen toteuttamista. (Antaki & Widdicombe 1998, 6; Lerner, Zimmerman & Kidwell 2011, 57; Levinson 2006, 46, 54.)

Tahmean medialaitteen ilmiössä perinteinen keskustelun vierusparijäsennys jollain tapaa häiriintyy. Tarkoitamme tahmean medialaitteen käsitteellä sitä, että älypuhelinta tai jotakin muuta ruutumedialaitetta käyttävä henkilö jää "kiinni laitteeseensa" eikä pysty osallistumaan samaan aikaan käynnissä olevaan keskusteluun täysipainoisesti. Voi olla, että häntä ei saa esimerkiksi vastaamaan kysymykseen. Tämä vaikuttaa siihen, miten muut keskustelijat vuorovaikutusta tulkitsevat ja ymmärtävät. (Mantere & Raudaskoski 2015; 2017.)

Tahmea medialaite on helposti arjesta tunnistettava ilmiö. Viittaamme käsitteellä tilanteisiin, jossa kasvokkainen keskustelu ja jonkin ruutumedialaitteen, yleensä älypuhelimen, käyttö ovat läsnä yhdessä ja samassa vuorovaikutuskontekstissa. Älypuhelimella kesken olevalla toiminnolla on vuorovaikutuksellista pakottavuutta, mikä tekee laitteesta hankalasti hylättävän. Vuorovaikutuksellisella pakottavuudella emme tarkoita ainoastaan sitä, että laitetta käytettäisiin välttämättä toisten ihmisten kanssa toimimiseen vaan viittaamme myös käyttäjän ja laitteen väliseen vuorovaikutukseen. Käyttäjän tehdessä jonkin toiminnon laitteella, laitteessa tapahtuu jokin tilan muutos, joka taas edellyttää uusia toimia käyttäjältä ja niin edelleen (vrt. Arminen 2005, 203).

Keskeistä medialaitteen tahmeudessa on se, että tilanteessa syntyy kaksi yhtäaikaista, kilpailevaa vuorovaikutuskontekstia. Kahden eri vuorovaikutuskontekstin – laitteen käytön ja kasvokkaisen keskustelun – limittyminen johtaa tavanomaista kasvokkaista vuorovaikutusta kompleksisempiin vuorovaikutuskäytäntöihin. Aiemmat analyysimme osoittavat, että toiminta älypuhelimen kanssa vaikuttaa samanaikaiseen keskusteluun muun muassa viipymisinä, takelteluina ja uudelleen aloittamisina (Mantere & Raudaskoski 2015; 2017). Nämä hankaloittavat kasvokkaisen vuorovaikutuksen tulkintaa. Emme siis tarkoita medialaitteen tahmeudella tilanteita, joissa yksilö on yksin uppoutunut mediasisältöihin eikä ole samanaikaisesti muussa vuorovaikutuskontekstissa mukana.

Käytämme seuraavaa aineistoesimerkkiä[1] taustoittamaan tahmean medialaitteen ilmiötä. Esimerkki havainnollistaa sitä, miten aina käsillä olevat puhelimet voivat keskeyttää meneillään olevan kasvokkaisen toiminnon. Keskustelun säännönmukaisuuteen nojaten tiettyyn vuorovaikutusaloitteeseen odotetaan tietynlaista vastausta tietyssä kohtaa vuorovaikutusta. Esimerkissä 7-vuotias tytär pyytää äitiään auttamaan häntä palapelin tekemisessä. Pyyntö on yksi perustavanlaatuinen ihmisten välisessä keskustelussa toteutuva toiminto, jonka perusrakenne opitaan jo varhaisessa lapsuudessa. Pyyntöön odotetaan tavallisesti jonkinlaista vastausta, ja kokonaan vastaamatta jättämistä pidetään yleensä epäkohteliaana tai outona. Sohvapöydän äärellä palapeliä kokoava tytär on toistuvasti pyytänyt äitiään osallistumaan palapelin rakentamiseen, mutta televisiota katseleva äiti ei ole suostunut. Juuri ennen nyt esitettävää aineistokatkelmaa tytär on osoittanut mieltään äidin kieltäytymisille liittäen dramaattisen isoeleisesti kahta palapelin osaa yhteen ja sanonut "ettei pysy edes koko palapeli yhdessä" (laajemman vuorovaikutusepisodin analyysi, ks. Mantere & Raudaskoski 2015, 217–224). Episodin lopussa tapahtuu vuorovaikutuksellisesti jotain mielenkiintoista, kun tyttären neljäs perättäinen pyyntö saada äidiltä apua saa vierusparikseen äidiltä erikoisen vastauksen.

    Esimerkki 1. (Suluissa olevat numerot ilmoittavat tauon pituutta sekunneissa)

    1 Tytär:  Voiks tulla auttaan.

    2 Äiti:    Mite sulla menee sielä.

    3 Tytär:  Missä.

    4        (1.0)

---

[1] Aineisto oli kerätty kodeista asettamalla videokameroita kuvaamaan kodin arkisia toimintoja. Aineiston keruu keskittyi medialaitteiden käyttöön osana Suomen Akatemian rahoittamaa projektia *Media, perheen vuorovaikutus ja lasten hyvinvointi* (johtajana Anja Riitta Lahikainen).

5　Äiti:　　Onko asiakkaita.

6　Tytär:　Oho.

7　　　　　(2.0)

8　Äiti:　　Joo.

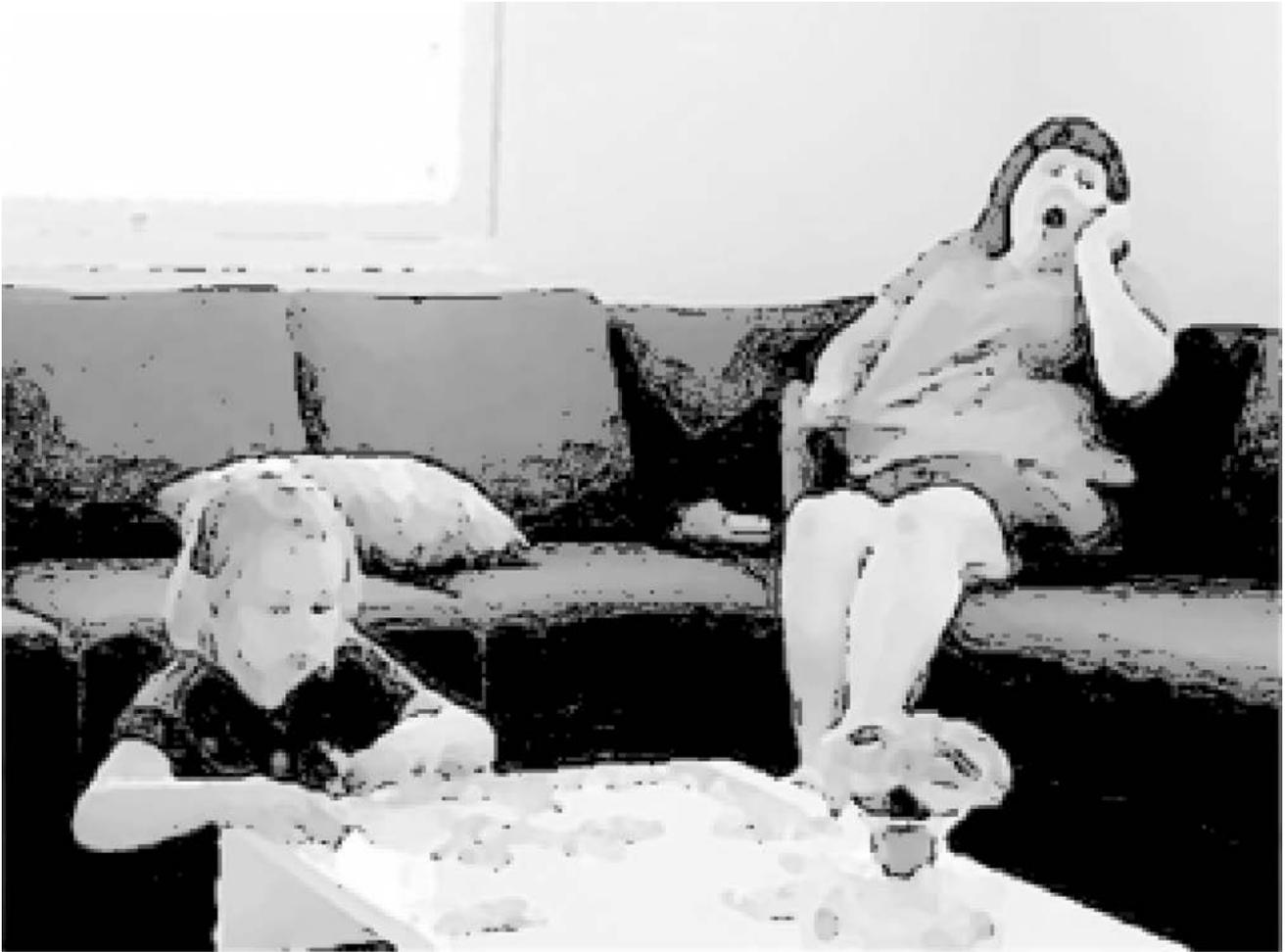

Kuva 1. Äiti aloittaa puhelinsoiton

Aineisto-otteen aikana tyttären katse on palapelissä ja äiti istuu viistosti hänen takanaan sohvalla. Niinpä tytär ei ole huomannut, että äiti on ottanut sohvalla vieressään olleen kännykän käteensä ja aloittanut puhelun (Kuva 1). Tyttären toistuvat avunpyynnöt äidille tulla auttamaan häntä palapelin kokoamisessa ja äidin kieltäytymiset ovat luoneet tietynlaisen yhteisen jatkuvan toiminnan kehyksen tyttären ja äidin välille. Myös heidän kehollinen sijoittumisensa tilaan, lähelle toisiaan, ja tässä asetelmassa yhdessä keskusteleminen jo pidemmän aikaa ovat vahvistaneet yhteisen toiminnallisen kehyksen olemassaoloa (vrt. Meyer, Streeck & Jordan 2017). Vaikka äiti on kieltäytynyt osallistumasta palapelin rakentamiseen, tyttären näkökulmasta televisiota katselevan äidin kanssa on ollut mahdollista virittää yhteinen osallistujakehikko, jossa pyyntöön suostumisen voi ottaa uudelleen esiin ja neuvoteltavaksi. Toisin sanoen televisio ei ole muuttunut sillä tapaa "tahmeaksi medialaitteeksi", että se olisi estänyt tai varsinaisesti hankaloittanut äidin ja tyttären välistä keskustelua.

Kun tytär neljännen kerran pyytää äitiä auttamaan palapelin tekemisessä, pyyntöä seuraakin äidin esittämä kysymys "Miten sulla menee siellä" (rivi 2). Äiti esittää kysymyksen juuri siinä kohdassa, jossa keskustelun jäsentyneisyyden perusteella olisi odotettavissa vastaus tyttären kysymykseen, mutta äidin vuoron sisältö on outo ja se saa tyttären pyytämään lisäselitystä: "Missä" (rivi 3). Seuraavaksi tytär kääntää katseensa kohti äitiä ja huomaa äidin puhuvan puhelimessa. Äidin vuoro "Onko asiakkaita" (rivi 5) on suunnattu puhelun keskustelukumppanille, ei tyttärelle, ja tytär ilmaiseekin tajuavansa tilanteen sanomalla "Oho" (rivi 6). Äiti jatkaa puhelinkeskustelua reagoimatta mitenkään tyttären pyyntöön ja hänen lyhyt "Joo"-lausumansa on osoitettu puhelun keskustelukumppanille.

Esimerkki kuvastaa, miten helposti saatavilla olevaan puhelimeen voidaan tarttua hetken mielijohteesta eikä omaa tekemistä välttämättä lainkaan selitetä muille läsnäolijoille. Tässä esimerkissä on kyse "perinteisestä" kännykän käytöstä eli puhelinsoiton aloittamisesta. Tilanne kuitenkin edustaa sitä yleistä piirrettä, että puhelimen käyttöä ei välttämättä pidetä selontekovelvollisuutta aiheuttavana edes silloin, kun se tapahtuu keskellä kasvokkaista vuorovaikutusta, jopa keskellä niin kutsuttua vierusparisekvenssiä (kuten pyyntö–vastaus).

Esimerkissä kännykän käyttö rikkoo äidin ja lapsen meneillään olevaa vuorovaikutusta. Kuitenkin havaittuaan äidin syventyneen puhelinkeskusteluun toisen kanssa, lapsi luopuu vaatimuksestaan saada apua palapelin kokoamisessa. Mielenkiintoista on, että televisiota katsovan äidin puheenvuorot vähitellen köyhtyvät ja yksipuolistuvat myös puhelinkeskustelun aikana. Näin ollen televisiokin alkaa näyttäytyä tahmeana medialaitteena suhteessa puhelinkeskusteluun.

Esimerkki havainnollistaa yleisemmin sitä, että kännykät osana kasvokkaisia keskustelutilanteita luovat uudenkaltaisia vuorovaikutuskonteksteja ja pakottavat keskusteluun osallistuvia soveltamaan muita kuin keskustelusta tuttuja vuorovaikutuksen sääntöjä näissä uusissa tilanteissa. Tämä käy ilmi myös nuorten keskinäisten vuorovaikutustilanteiden analyysistämme, jota esittelemme artikkelissa tuonnempana. Edellisessä osiossa käsittelemämme saavutettavuuden ja saatavilla olon käytänteet toivat esiin, että kun puhelinta käytetään kesken meneillään olevaa muuta vuorovaikutusta, se vaikuttaa väistämättä jollain tapaa myös kasvokkaisiin vuorovaikutuskonteksteihin.

Olemme aiemmissa tutkimusjulkaisuissa (Mantere & Raudaskoski 2015; 2017) kuvailleet keskustelunanalyysiä hyödyntäen medialaitteen tahmeutta tilanteissa, jossa puhelimen käyttäjän on hankala suunnata huomiotaan paikalla olevaan keskustelukumppaniin. Olemme esittäneet analyysin tilanteesta, jossa 12-vuotias tytär pyytää äitiään katsomaan piirtämiään kuvia. Kun hän ei saa sanallisesti äidin katsetta irtoamaan kännykän ruudulta, hän lisää puhuttuihin pyyntöihin myös kehollisia elementtejä. Tytär esimerkiksi naputtelee ensin äitiä olkapäähän ja tökkää sitten äitiä käsivarteen saadakseen hänet lopulta luopumaan puhelimesta.

Älypuhelinta käyttävän henkilön tarkkaavaisuus on väkisinkin keskittynyt parhaillaan puhelimen kanssa suoritettavaan toimintoon, eikä siitä irrottautuminen välttämättä tapahdu silmänräpäyksessä, varsinkin jos tehtävä on kesken. Älypuhelimella toimiminen vaatii sekä retrospektiivistä että prospektiivista toiminnan arviointia (vrt. Reed 1996, 12–19). Mitä erilaisempia toimintoja mahdollistavien älylaitteiden käyttötilanteet eivät toisin sanoen ole rutiininomaisia vaan vaativat käyttäjän tilannesidonnaista arviointia ja toimimista tuon arvion mukaan. Esimerkiksi viestiin voi joku olla odottamassa vastausta taikka pankkiyhteys voi katketa, jos asiaa ei hoida riittävän nopeasti loppuun. Kuitenkin samalla myös kasvokkainen keskustelu vaatii jatkuvaa tilanteen arvioimista: pysyäkseen keskustelussa "kärryillä" ja pystyäkseen tekemään ymmärrettäviä puhetekoja, henkilön tulee olla tietoinen siitä, mistä on puhuttu ja mitä omasta puheenvuorosta seuraa (vrt. Garfinkel 1967, 41).

Myös psykologisissa tutkimuksissa on havaittu, että sekventiaalisesti, vaihe vaiheelta, etenevissä ja siten jatkuvaa harkintaa vaativissa toiminnoissa kaksi yhtäaikaisesti tietoista arviointia vaativaa tehtävää synnyttävät "kognitiivisen pullonkaulan". Tästä seuraa suorittamistehossa havaittava lasku. (Pashler 1994; Pashler & Johston 1998.) Nuorille ja nuorille aikuisille suoritetuissa laboratoriokokeissa eri tehtävien yhtäaikaisen tekemisen on todettu vaikuttavan suoritustehoon ja kuormittavan tarkkavaisuutta sääteleviä aivoalueita siitä huolimatta, että koehenkilöt ovat haastatteluissa kertoneet, ettei usean eri median tai sovelluksen yhtäaikainen käyttäminen (*media multitasking*) vaikuta heidän suoritustehoonsa (Moisala ym. 2016). On kuitenkin niin, ettei kahta tietoista arviointia vaativaa toimintoa voida suorittaa täsmälleen yhtä aikaa. Sen sijaan yhtäaikaisia toimintoja synkronoidaan lomittain, mikä vaikuttaa molempiin meneillään oleviin toimintoihin. (Bowman ym. 2010; Levy & Gardner 2012.) Keskustelutilanteessa yhtäaikaisten toimintojen suorittaminen voi näyttäytyä nimenomaan tahmean medialaitteen ilmiöön liittyvinä taukoina ja takelteluina.

## Sivustakatsojan pimento

Kasvokkaiseen vuorovaikutukseen osallistuva harvoin tietää, mitä samassa tilanteessa älypuhelintaan käyttävä puhelimellaan tekee, ellei hän puhu puhelimessa, kuten aiemmin esittämässämme esimerkissä. Älypuhelimen käyttäjä sijoittaa tavallisesti puhelimen ruudun kohden kasvojaan, eikä sivustakatsojalle muodostu tällöin mahdollisuutta nähdä, mitä laitteella tehdään. Tämä on yksi keskeinen elementti, joka on vaikuttanut kasvokkaisen vuorovaikutuksen uudenlaiseen jäsentymiseen älypuhelinten aikakaudella, myös tahmean medialaitteen kaltaisissa vuorovaikutustilanteissa. Tässä osiossa pohdimme tarkemmin tätä sivustakatsojan pimennoksi nimeämäämme ilmiötä. Olemme aikaisemmin esittäneet siitä teoreettisen mallinnuksen (Raudaskoski ym. 2017). Tässä artikkelissa esittelemme myös sivustakatsojan pimentoon liittyvän kokeellisen tutkimuksemme tuloksia. Selitämme tutkimusasetelmaa ja tuloksia tarkemmin kuvattuamme ensin sivustakatsojan pimennon käsitteen perusperiaatteita.

Sivustakatsojan pimento liittyy tahmean medialaitteen tavoin tilanteisiin, joissa samassa fyysisessä tilassa olevien vuorovaikutuksen osapuolten "väliin" älypuhelimen käyttö asettuu. Koska mobiili älypuhelin on henkilökohtainen ja kulkee mukana kaikkialle, vuorovaikutus laitteen kanssa voi sijoittua keskelle melkein mitä tahansa muuta vuorovaikutuskontekstia, varsinkin kun laite voi myös kutsua (äänen, värinän yms. avulla) kesken kaiken omistajaansa tähän toiseen vuorovaikutuskontekstiin (vrt. Radesky ym. 2016, 699).

Henkilökohtaisuuteen liittyy myös yksityisyys, jota tukevat sekä laitteen tekniset ominaisuudet että laitteen käyttöön liittyvät sosiaaliset normistot. Älypuhelimissa on tavallisesti kosketusnäyttö, jota katsellaan ja jonka kautta laitetta myös käsitellään. Näyttö sijoittuu yleensä suhteessa käyttäjän kasvoja niin, että sivustakatsojalle harvoin paljastuu näytön näkymä tai se, mitä käyttäjä sormillaan koskettaa. Teknisten ominaisuuksien lisäksi myös sosiaaliset normit tukevat yksityisyyttä: toisen henkilön älypuhelimen ruudun sisältö kuuluu yksityisyyden piiriin ja yleensä sen tarkastelua tulisi välttää, ellei sisältöä erityisesti jaeta muiden kanssa (Raudaskoski ym. 2017, 180).

Lisäksi älypuhelin mahdollistaa lukemattomien erilaisten toimintojen suorittamisen alati lisääntyvien sovellusten avulla. Kuitenkin sen toimimisesta puuttuvat sellaiset keskeiset toiminnan tunnistamista helpottavat vihjeet, joita muissa sosiaalisessa ympäristössä tehdyissä toiminnoissa yleensä on ja joista sivustakatsoja voi tehdä arvioita meneillään olevasta toiminnasta. Henkilö, joka on kasvokkaisessa toimintatilanteessa mukana, tai haluaisi ehkä aloittaa vuorovaikutuksen älypuhelinta käyttävän henkilön kanssa, saa poikkeuksellisen vähän vihjeitä siitä, mitä laitetta käyttävä on tekemässä, missä vaiheessa toiminto on ja onko sitä soveliasta keskeyttää (Mantere ym. 2018).

Yksi sivustakatsojan pimennon käsitteen taustalla olevista teorioista on ekologisen psykologian ajatus tarjoumista (Gibson 1979). Tarjoumateorian lähtökohtana on ajatus siitä, että fyysinen ympäristö sekä toiset ihmiset tarjoavat meille erilaisia toiminnan mahdollisuuksia. Esimerkiksi esineet mahdollistavat monenlaisia käyttötapoja, ja muun muassa lapset tai vaikkapa jotakin ammattitaitoa opettelevat henkilöt voivat oppia sivusta katsomalla, miten ja millaisiin toiminnan päämääriin toiset ihmiset esineitä käyttävät (Ingold 2000, 22). Michael Tomasello kutsuu tätä prosessia esineiden intentionaalisten tarjoumien oppimiseksi (Tomasello 1999a, 84; 1999b, 165–166).

Aiemmassa tutkimuksessa (Raudaskoski ym. 2017) tekemämme teoreettisen mallinnuksen tulos oli, että kaikista nykykodeista löytyvistä yksittäisistä artefakteista älypuhelin mahdollistaa eniten erilaisia toimintoja. Toisin sanoen sillä on lukuisia erilaisia tarjoumia. Samanaikaisesti älypuhelimen käyttö antaa sivustakatsojalle vähiten vihjeitä siitä, mihin toimintoon laitetta käytetään. Niinpä älylaitteen intentionaaliset tarjoumat pysyvät sivustakatsojalle piilossa.

Halusimme testata sivustakatsojan pimennon käsitettä myös empiirisesti. Rakensimme etnometodologiaa, eläytymismenetelmää ja perinteistä kyselytutkimusta soveltavan koeasetelman, johon osallistui 112 yliopisto-opiskelijaa. Poimimme aiemman tutkimusprojektimme naturalistisesta videoaineistosta[2] otteita, joissa mieshenkilö istuu nojatuolissa ja käyttää älypuhelinta tai lukee lehteä samaan aikaan, kun kuvan ulkopuolelle jäävä henkilö pyytää häntä keskustelemaan kanssaan saamatta vastausta. Piirsimme tilanteiden pohjalta kaksi sarjakuvaa, joista teimme keskenään identtiset lukuun ottamatta henkilön käyttämän mediavälineen muutosta. Ensimmäisessä sarjakuvassa Mattina puhuteltu mies käyttää älypuhelinta ja toisessa hän lukee lehteä tai kirjaa. Matille kysymyksiä esittävän henkilön identiteetti jätettiin tarkoituksella piiloon, koska halusimme tutkimukseen osallistuvan eläytyvän nimenomaan kysymyksen tekijän rooliin.

---

[2] Käytimme samaa videoaineistokokoelmaa kuin Esimerkissä 1.

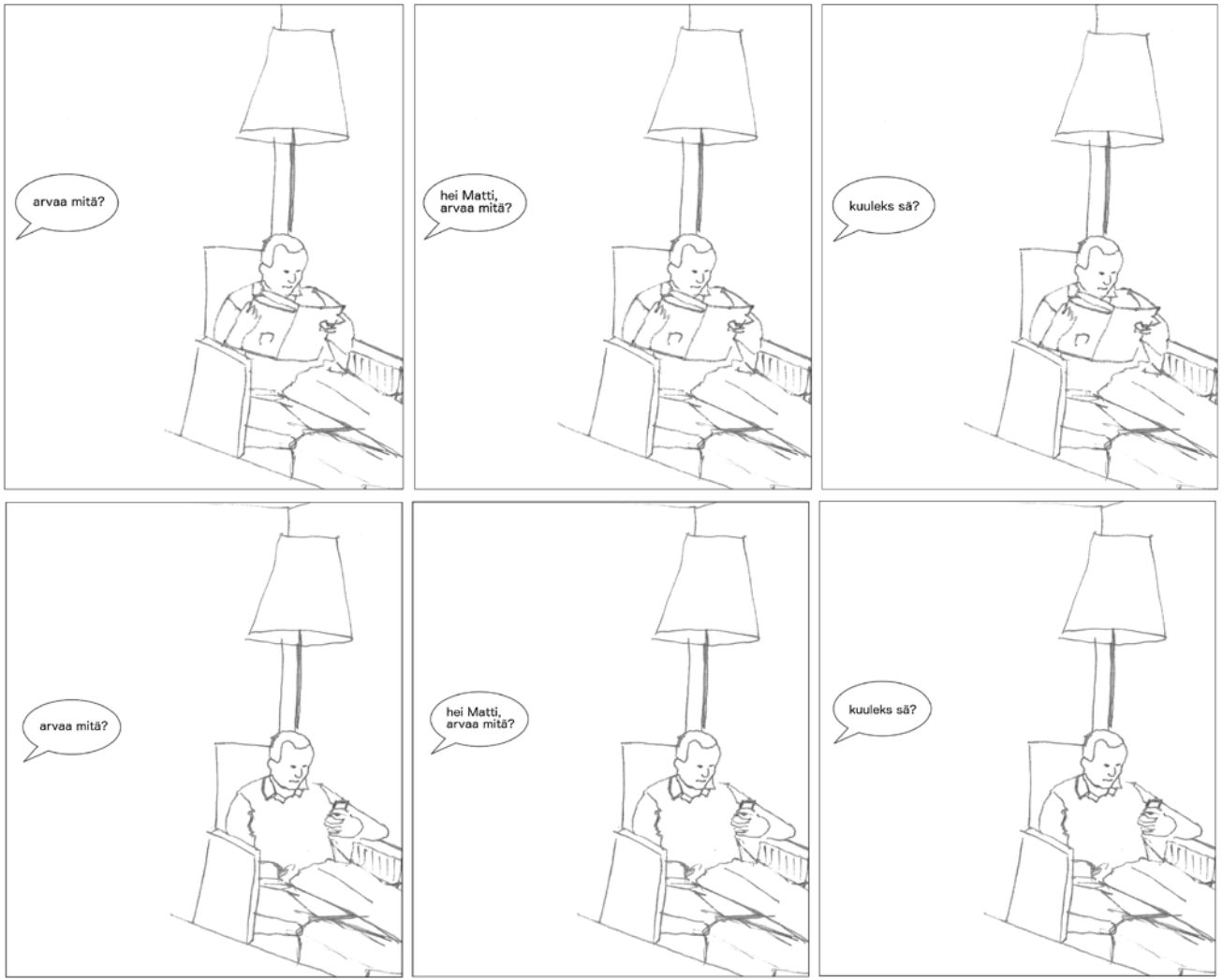

Kuva 2. Verkkokyselyn sarjakuvat

Teimme sarjakuvista kyselylomakkeen internetiin. Kyselyn hypoteesina oli, että vastausta vaille jäävä vuorovaikutusaloite ärsyttää ihmisiä. Hypoteesimme perustui Harold Garfinkelin (1963) rikkomuskokeiden tuloksiin, joissa vakiintuneiden vuorovaikutussääntöjen, kuten kysymys–vastaus-vuoroparin, rikkomisen todettiin aiheuttavan usein toisissa osallistujissa suuttumusta ja ärsyyntyneisyyttä. Varioimalla tilannetta halusimme selvittää, vaikuttaako havaittavissa oleva vastaamattomuuden syy (puhelimen käyttö tai kirjan tai lehden lukeminen) arviointeihin. Pyytämällä avoimia vastauksia selvitimme myös sitä, millaisia sanallisia selontekoja ihmiset itse antavat reaktioilleen.

Pyysimme vastaajia katsomaan sarjakuvaa, eläytymään siinä puhuvan henkilön asemaan ja arvioimaan Likert-asteikolla yhdestä viiteen[3] kuinka ärsyttäväksi he tilanteen kokevat. Vastaaja tuli tietoiseksi toisesta tutkimukseen kuuluvasta sarjakuvasta vasta annettuaan vastauksen ensimmäiseen. Jaoimme vastaajat kahteen ryhmään niin, että 56 sai ensin nähtäväkseen älypuhelin-sarjakuvan ja toiseksi lehti-sarjakuvan. Toisille 56 vastaajalle järjestys oli käänteinen. Kun vastaaja oli arvioinut myös toisen sarjakuvan ärsyttävyyden, pyysimme häntä omin sanoin kertomaan, minkä vuoksi hän arvioi ensimmäisen ja toisen sarjakuvan kuten arvioi. Viisi vastausta poistettiin puuttuvien vastausten tai kirjallisesta vastauksesta ilmi käyneen tehtävän väärin ymmärtämisen vuoksi.

Kukaan kyselyyn vastanneista ei pitänyt lehden lukemisen johdosta vastaamatta jättämistä ärsyttävämpänä kuin vastaamattomuutta kännykän käytön takia. "Älypuhelimen"[4] käytön takia puuttuvan vastauksen kokivat ärsyttävämpänä 52,3 prosenttia vastaajista ja 47,7 prosenttia pitivät molempia tilanteita yhtä ärsyttävinä ($\chi^2$(2)=53.850, p<0,001, n=107). Keskiarvoissa tämä näkyi niin, että vaikka molemmat tilanteet arvioitiinkin ärsyttäviksi, niin kännykän käyttö ärsytti enemmän (lehden luku ka=3,13, kh=0,946; kännykän käyttö ka=3,87, kh=0,870; n=107, p<0.001). Puhelimen käytön suurempi ärsyttävyys suhteessa lehden lukuun sai vielä hienoisen vahvistuksen siitä, jos puhelimen käyttötilanne arvioitiin ensin (lehti-tilanne ensin: +0,585, n=53; kännykkä-tilanne ensin: +0,889, n=54; p<0,001). Vastauksia voidaan tulkita niin, että 1) vastaamattomuus sinällään ärsyttää, ja 2) älypuhelimeen uppoutuminen vastaamattomuuden syynä ärsyttää enemmän kuin kirjan tai lehden lukeminen.

---

[3] 1 = En lainkaan ärsyttäväksi, 2 = Melko vähän ärsyttäväksi, 3 = Jonkin verran ärsyttäväksi, 4 = Melko paljon ärsyttäväksi, 5 = Erittäin ärsyttäväksi

[4] Piirroskuvasta ei voi nähdä, onko Matin kädessään pitämä matkapuhelin internetominaisuuksilla varustettu älypuhelin, mutta se oli oletuksena, koska hän nimenomaan katsoi ruutua.

Kun kännykän käyttö ärsytti enemmän, osallistujat selittivät tätä muun muassa sillä, etteivät voi tietää, mitä henkilö kännykällään tekee. Lehteä lukevan henkilön vastaamattomuutta tulkittiin kyselyvastauksissa näkyvillä olleen toiminnan kohteen avulla, ja siksi tilanne ei ärsyttänyt heitä yhtä paljon:

> Kännykän käyttö siis ärsyttää myös siksi enemmän, että en tiedä mitä henkilö kännykällään tekee, kun taas lehteä lukevasta henkilöstä näkee suoraan, miksi hän ei vastaa.

Kyselyn vastaajat siis nostivat merkitykselliseksi saman asian, mitä sivustakatsojan pimennon käsitteemme korostaa: kirjaa tai lehteä pitelevästä ja sitä katsovasta henkilöstä tietää, mitä hän tekee, mutta kännykkään orientoituneesta ei välttämättä tiedä, mitä tiettyä toimintaa hän on puhelimellaan suorittamassa ja miten kauan toiminto kestää.

Garfinkel on käyttänyt dokumentaarisen tulkintamenetelmän käsitettä kuvatessaan sitä mekanismia, jonka avulla ihmiset jatkuvasti ja tiedostamattaankin tulkitsevat eri tilanteiden mieltä. Tilanteissa esiin tulevien yksittäisten havaintojen – dokumenttien – avulla ihmiset tekevät päätelmiä siitä, "mitä nyt on meneillään". Tehtyään tietynlaisen tulkinnan tilanteen mielestä, he tarkastelevat muitakin tilanteessa esiin tulevia asioita tämän kehyksen kautta. Näkyvät evidenssit, dokumentit, siis samanaikaisesti sekä auttavat muodostamaan toiminnalle tietyn tulkintakehyksen että tulevat tulkituksi tämän oletetun kehyksen kautta. (Garfinkel 1967; Heritage 1984, 84–86.) Tekemässämme tutkimuksessa kävi ilmi, että sivustakatsojan pimennon takia älypuhelinta käyttävän toiminnan dokumentaarinen tulkinta hankaloituu. Vastaavasti niissä tapauksissa, joissa on esillä tarpeeksi dokumentteja toiminnan kohteesta, kuten lehden tai kirjan lukemisessa on, vastaamattomuudelle voidaan antaa tulkintoja, jotka ovat enemmän ymmärrettäviä ja vähemmän ärsyttäviä. Tätä kuvastaa myös erään tutkimukseen osallistuneen vastaus:

> Ensimmäinen tilanne [kännykän käyttö] on todella ärsyttävä, koska en tiedä, mikä on paljo tärkeämpää kuin asiani. Toisessa tilanteessa on mahdollisesti joku hyvä kirja tai läksy kirja, jonka tilanteen ymmärrän, että toinen haluaa keskittyä.

## Tahmea medialaite ja sivustakatsojan pimento: normaalistuvat käytänteet ja älypuhelinajan kasvotyö

Tunnistettuamme, nimettyämme ja empiirisesti todennettuamme tahmean medialaitteen ja sivustakatsojan pimennon ilmiöt, olemme keränneet uutta tutkimusaineistoa älypuhelimen käytön vaikutuksista nuorten ja nuorten aikuisten keskusteluihin (naturalistinen videoaineisto)[5] sekä heidän mielipiteistään puhelimen käytöstä (eläytymismenetelmä)[6]. Olemme analysoineet naturalistista videoaineistoa keskustelunanalyyttisin metodein. Tämän artikkelin tutkimustehtävän kannalta videoaineistosta nouseva keskeinen analyyttinen huomio on se, että tahmean medialaitteen ja sivustakatsojan pimennon ilmiöitä ei nuorten aktiivisesti älypuhelinta käyttävien keskuudessa sanallisesti problematisoida, vaan käytännössä ne hyväksytään osaksi kasvokkaista keskustelua. Tuomme jäljempänä kuitenkin esiin, että vaikka toisia ei huomautetakaan laitteen käytöstä, se saattaa silti herättää tunteita toisissa keskustelijoissa.

Vuorovaikutustilanteissa on erilaisia strategioita, joilla pysytellään keskustelussa mukana, vaikka toimitaankin samanaikaisesti puhelimen kanssa. Katsekontakti ja kehon asento ovat eräitä olennaisia keinoja orientoitumisen osoittamiseksi keskustelun aiheisiin, vaiheisiin ja keskustelijoihin, mutta kysymys ei ole pelkästään niistä. "Tahmeaan laitteeseen" uppoutunutkin voi tietyillä tavoilla ilmaista osallisuuttaan meneillään olevassa kasvokkaisessa vuorovaikutuksessa ilman, että hän siirtää orientaatiotaan pois älypuhelimesta. Seuraavassa tekstilitteraatiossa käy ilmi, miten kolmen nuoren henkilön kahvilakeskustelussa Henkilö 2 (H2) pitää itseään mukana kasvokkaisen vuorovaikutuksen osallistumiskehikossa, vaikka hän koko ajan samanaikaisesti katsoo puhelintaan ja kirjoittaa sillä jotakin. Litteraation alla oleva kuva havainnollistaa tilannetta.

---

[5] Naturalistisia videoaineistoja monenkeskisistä keskusteluista: seitsemän kahvi/ruokapöytäkeskustelua, jotka ovat kestoltaan puolesta tunnista kahteen tuntiin ja joissa oli osallistujia kolmesta viiteen.

[6] Eläytyminen videomateriaaliin: 119 osallistujaa.

Esimerkki 2. (Hakasulut kuvaavat päällepuhunnan alkamisen paikkoja, (.) pientä taukoa sekä (h) naurua puheen seassa)

1 H1:    Mua ärsyttää kun äiti ja iskä rupee (.) aina kun tulee joku juominen

2         puheeksi niinku just silloin kun mä olin lähössä sinne potkiaisiin ja

3         määä täy[ttelin sitä mun taskumattia

4 H2:            [Joo.

5 H1:    k<u>ei</u>[ttiös(h)sä.

6 H2:        [Joo.

7 H3:    Niinku enää se ei [onneks oo niin semmone juttu,

8 H2:                      [Joo.

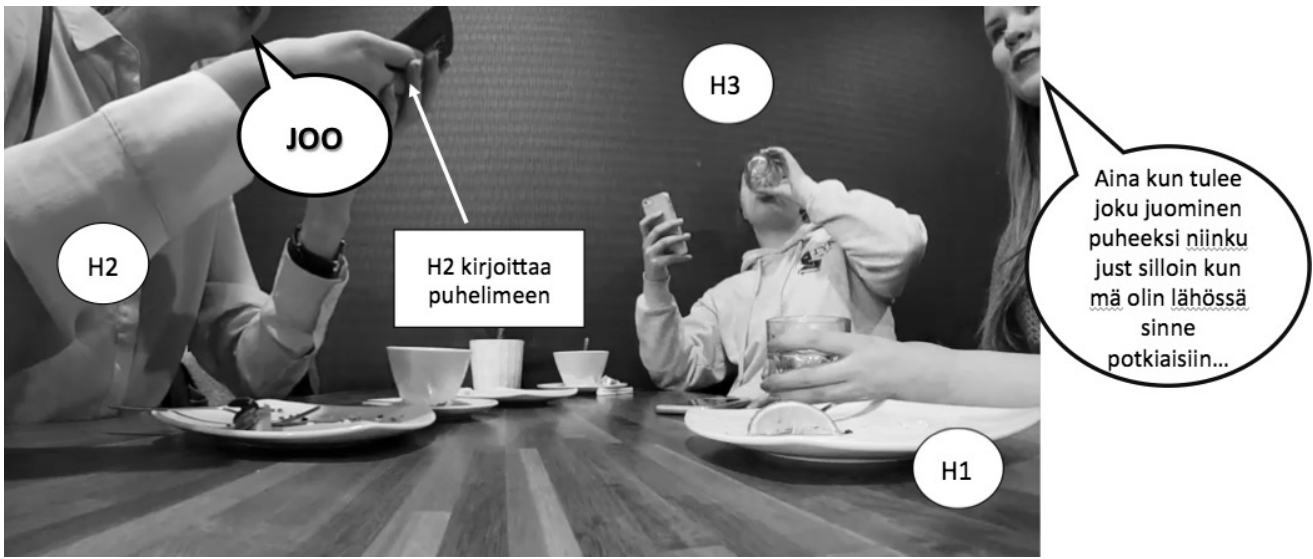

Kuva 3. Laitetta käyttävä pysyy minimipalautteiden avulla keskustelussa mukana

Suomen kielessä dialogipartikkelin "joo" käyttäminen nähdään usein joko kuulijaksi asettautumisena, samanmielisyyden osoittamisena tai keskustelusekvenssin päättämisenä (Sorjonen 2001). Esimerkissämme "joo" osoitti erityisesti kuulolla oloa, sillä Henkilö 2 kirjoitti koko ajan samanaikaisesti jotain älypuhelimellaan peukaloitaan käyttäen. Samalla esimerkistä voi nähdä medialaitteen tahmeuden, kun joo-partikkelit eivät osukaan "siirtymän mahdollistaviin kohtiin", joiden yhteydessä ne yleensä esiintyvät. Tavallisesti partikkelit ovat siis sellaisissa kohdissa, joissa puheenvuoro voi vaihtua toiselle henkilölle ja joita yleensä edeltää pieni tauko, mutta tässä esimerkissä Henkilö 2 tuottaa joo-partikkelit "ohi" niiden tyypillisistä esiintymiskohdista. Tämän voi nähdä ilmentävän sitä haasteellisuutta, joka seuraa vuorovaikutuksesta kahdessa eri temporaalisessa järjestyksessä: samaan aikaan kun Henkilö 2 pitää joo-lausumillaan yllä osallisuuttaan keskusteluun, hän ei kuitenkaan aivan kokonaan ole tämän osallistumiskehikon käytettävissä.

Aineistossa on edellisen esimerkin kaltaisten tilanteiden lisäksi episodeja, joissa nuoret käyttävät laitteitaan kesken kasvokkaisen keskustelun ilman minkäänlaisia selityksiä muille läsnä oleville. Jos puhelimen käyttäjän katsekontakti toisiin sekä Esimerkissä 2 kuvatun kaltaiset sanalliset minimipalautteet puuttuvat, ainoa kasvokkaisen vuorovaikutuksen kannalta merkityksellinen seikka, jonka hän vuorovaikutuskontekstiin tuo on se, että toiset voivat *nähdä* hänen olevan uppoutunut puhelimeensa.

Itse asiassa laitteen käyttäjä voi nojata tähän samassa fyysisessä tilassa olevien muiden henkilöiden tekemään dokumentaariseen tulkintatyöhön: toiset näkevät, että hän käyttää älypuhelintaan ja he voivat siten sopeuttaa vuorovaikutustekonsa siihen. Sivustakatsojan pimennosta on siis tullut älypuhelinajan väistämätön sekä tietyllä tapaa jopa oletuksenmukainen vuorovaikutuksellinen positio. Vaikka sivustakatsoja ei tietäisikään, mitä laitetta käyttävä tarkalleen tekee, hän kuitenkin tietää, että käyttäjä orientoituu laitteeseen.

Sen sijaan se taho, jonka kanssa käyttäjä on älypuhelimen kautta vuorovaikutuksessa, ei yleensä voi tehdä päätelmiä siitä kehollisesta toimintaympäristöstä, jossa laitteen käyttäjä samanaikaisesti on osallisena. Niinpä älypuhelimen käyttäminen voi tietyissä tilanteissa olla vuorovaikutuksellisesti jopa pakottavampaa kuin kasvokkainen keskustelu. Läsnä olevat tietävät, että puhelin pidättelee, mutta "puhelimen toisessa päässä oleva" ei välttämättä tiedä, että kasvokkaiset keskustelijat vaativat osallistumista keskusteluun. Tämä pätee myös tilanteissa, joissa puhelimen käyttäjä ei suoranaisesti kommunikoi kenenkään kanssa vaan käyttää jotain tietoteknistä sovellusta: kuten aiemmin toimme esiin, monet sovellukset vaativat käyttäjän etenevän vaihe vaiheelta, eivätkä nekään "pysty" huomioimaan, missä muussa mahdollisessa vuorovaikutustilanteessa käyttäjä on.

Olemme tehneet huomion, että nuorten kasvokkaisissa vuorovaikutustilanteissa älypuhelimen käyttö ja tahmean medialaitteen sekä sivustakatsojan pimennon kaltaiset ilmiöt ovat sallittuja. Tämä on suuri muutos traditionaalisiin keskustelun sääntöihin verrattuna. Vuorovaikutustilanteessa nuoret luottavat siihen, että kehollisessa toimintaympäristössä olevat 1) näkevät että käyttäjä orientoituu älypuhelimeen, 2) sallivat tämän ja 3) kaikki osallistujat jakavat kohdat 1) ja 2). Osallistujat ovat siis Harold Garfinkelin määritelmää seuraten tietoisia uudesta normista ja sen rikkomiseen liittyvästä selontekovelvollisuudesta. (Vrt. Garfinkel 1963, 209; Heritage 1984, 117.)

Olemme nimenneet tämän uuden normin "älypuhelinajan kasvotyöksi". Tarkoitamme sillä sitä, että osallistujat jakavat säännön, jonka mukaan älypuhelimen käyttö kesken keskustelun sallitaan. Älypuhelinajan kasvotyö liittyy tahmean medialaitteen ja sivustakatsojan pimennon ilmiöihin, joihin mahdollisesti liittyvää vuorovaikutuksellista problematiikkaa uusi yhteisesti jaettu normi helpottaa. Tällöin muut keskustelijat eivät niin sanotusti tee numeroa laitteen käytöstä ja suojelevat näin toiminnallaan älypuhelimen käyttäjän kasvoja (vrt. Goffman 1955). Normin vastaista ei siis ole laitteen käyttäminen kesken keskustelun vaan laitteen käytön paheksuminen tai sen osoittaminen jonkinlaiseksi ongelmaksi.

Normin olemassaolo ei kuitenkaan poista sitä, etteikö perinteisen keskustelun rikkoontuminen synnyttäisi tunteita. Esimerkiksi kysymys–vastaus-vieruspari on vahvasti moraalisesti velvoittava rakenne, jolloin toisen uppoutuminen älypuhelimeen saattaa herättää ärsyyntymistä siitä huolimatta, että sitä ei kohteliaisuussyistä ilmaista avoimesti. Aiemmin esittelemämme "ärsyttävyyskyselymme" vastauksista on tulkittavissa, että moraalinen närkästys vastaamattomuudesta ei ole kadonnut mihinkään, varsinkaan jos vastaamattomuus johtuu älypuhelimen käytöstä.

Lisäksi toisessa aineistossa pyysimme tutkimukseen osallistujia (119 henkilöä) katsomaan ja kommentoimaan neljää videopätkää, joissa seurataan erään 14-vuotiaan tytön päivän tapahtumia kotona ja kavereiden kanssa.[7] Tulokset tukevat näkemystä, että älypuhelimen jatkuva näprääminen on normalisoitunut "nuorisokulttuuriksi", joka silti aiheuttaa nuorissa ambivalentteja tunteita vuorovaikutuksen ulkopuolelle jäämisestä. Esimerkiksi yhdessä videoklipissä kohdetyttö on bussipysäkillä kahden kaverinsa kanssa. Kaikilla on älypuhelimet esillä, mutta he samalla myös keskustelevat keskenään. Tätä katkelmaa kommentoidessaan tutkimukseen osallistujat tuovat esiin muun muassa sen, että videolla nähty vuorovaikutus on "tyypillistä nykynuorten käyttäytymistä".

---

[7] Aineiston on kuvannut dokumenttiohjaaja Ditte Uljas.

> Videolla tapahtui nähdäkseni melko tyypillistä nykynuorten käyttäytymistä, välillä medialaitteita vilkuillen ja selaillen sekä välillä läsnä keskustellen. […] Jos olisin änkyrä, sanoisin, että olisi mukava, jos seurassa ei näprättäisi kännykkää ja keskityttäisiin siihen, mutta se toisaalta olisi nykyisen nuorisokulttuurin sivuuttamista. (Nainen, 28 v.)

Sivustakatsojan pimento voi aiheuttaa sivullisuuden tunnetta, vaikka älypuhelinten näpräämistä pidettäisiinkin tavallisena. Vuorovaikutuksen sosiaalinen konteksti määrittää sitä, koetaanko sivulliseksi jäämisen tunne soveliaaksi ilmaista ääneen.

> Tunnistan videon tilanteen omasta elämästäni. Kaveriporukassani jotkut viettävät suurimman osan yhteisestä ajastamme kännykällä, joten he tuntuvat olevan enemmän emotionaalisesti läsnä jossain muualla, vaikka he olisivatkin fyysisesti lähellä. Koen itse jatkuvasti ärtymystä siitä, jos esimerkiksi ihmiset lounastaessaan kanssani ovat jatkuvasti kännykällä ja päivittävät sosiaalista mediaa. Se saa tuntemaan itseni tylsäksi ja epäkiinnostavaksi ihmiseksi. (Mies, 24 v.)

## Lopuksi

Olemme käsitelleet tässä artikkelissa älypuhelinajalla kasvokkaisessa keskustelussa esiintyviä uudenlaisia vuorovaikutusilmiöitä. Totesimme, että älypuhelin tarjoaa sivustakatsojalle hyvin vähän vihjeitä meneillään olevan toiminnon luonteesta, vaikka toisaalta tarjoaakin käyttäjälleen kaikista kodin artefakteista eniten erilaisia toiminnan mahdollisuuksia. Olemme käsitteellistäneet tämän ilmiön sivustakatsojan pimennoksi. Se eroaa merkittävästi toisten ihmisten muiden arkisten toimintojen, kuten lehden lukemisen tai vaikkapa ruuanlaiton, havainnoimisesta ja näiden toimintojen luonteen ymmärtämisestä.

Älypuhelin on erityinen myös sillä tavoin, että useat sillä suoritetut toiminnot ovat vuorovaikutteisesti jäsentyneitä niin, että joko toinen henkilö tai sovellus odottaa puhelimen käyttäjältä reagointia. Älypuhelimen toimintoihin voidaan uppoutua myös ilman tätä vaadetta. Monien toimintojen älypuhelin tuleekin helposti "tahmeaksi" ja sen käyttäminen hankalasti lopetettavaksi, vaikka kehollisessa toimintaympäristössä toiset ihmiset odottaisivat laitteen käyttäjän reagointia.

Älypuhelimet eroavat muista arkisista välineistä myös siinä, että ne voivat aktiivisesti kutsua henkilökohtaiseen vuorovaikutukseen: puhelin soi ja viestiäänet kutsuvat juuri tiettyä henkilöä, missä ja milloin tahansa. Etenkin aktiivisesti älypuhelinta käyttävät nuoret reagoivat nopeasti näihin kutsuihin, jopa niin, että ne ohittavat – ja niiden sallitaan ohittavan – meneillään olevan kasvokkaisen keskustelun.

Tutkimamme ja nimeämämme ilmiöt – tahmea medialaite ja sivustakatsojan pimento – ovat vuorovaikutuksellisia ilmiöitä, jotka liittyvät sellaisiin ruutumedialaitteiden käyttötilanteisiin, joissa kaksi henkilöä tai useampi henkilö ovat samassa fyysisessä toimintaympäristössä. Ne voivat siis ilmetä missä tahansa tilanteessa, jossa älypuhelimen käyttö on mukana kasvokkaisessa keskustelussa. Ilmiöinä ne kuvastavat tiettyjä kommunikatiivisia tarjoumia ja vuorovaikutusstrategioita, eivätkä sinällään ole sidoksissa esimerkiksi keskustelijoiden ikään, sukupuoleen tai johonkin muuhun piirteeseen. Sen sijaan ne ovat sidoksissa erilaisiin keskustelua ylläpitäviin jäsennyksiin, kuten vieruspari-, preferenssi- tai korjausjäsennyksiin. Tahmean medialaitteen ja sivustakatsojan pimennon analysoiminen ei ainoastaan täytä tärkeää tutkimuksellista aukkoa kasvokkaisen vuorovaikutuksen tutkimuksen kentällä vaan auttaa myös ihmisiä jäsentämään arkisia toimintojaan, joita älypuhelimen ja muiden mobiilien ruutumedialaitteiden käyttö on viime vuosina saattanut monimutkaistaa.

Videoaineistoihin perustuvien tutkimustemme mukaan nuorten keskinäisissä keskusteluissa tahmean medialaitteen ja sivustakatsojan pimennon kaltaisia ilmiöitä esiintyy paljon, mutta niitä ei erityisesti problematisoida, silloinkaan kun ne vaikeuttavat keskustelun sujuvuutta. Nuoret ovat omaksuneet erilaisia strategioita olla läsnä kahdessa limittyvässä osallistumiskehikossa, kuten vaikkapa käyttämällä kuulolla oloa ilmaisevaa joo-partikkelia. Sivustakatsojan pimentoa ei itse keskustelutilanteissa käsitellä ongelmallisena vuorovaikutuspositiona, vaan sivustakatsojan oletetaan ymmärtävän laitetta käyttävän positiota ja laitteen tahmeutta.

Tämän voisi ajatella olevan todiste siitä, että laitteisiin ja perustuvanlaatuisiin keskustelukäytänteiden muutoksiin ollaan jo totuttu. Kuitenkin muut, kyselyihin sekä virikemateriaaleihin pohjaavat kulttuurisia kokemuksia sanallistavat tutkimusaineistomme tuovat esiin, että vaikka uudet ilmiöt käytännön tasolla onkin hyväksytty osaksi normatiivista nuorisokulttuuria, ne kuitenkin saattavat herättää vuorovaikutustilanteissa ärtymystä ja erilaisia ulos jäämisen kokemuksia – aivan kuten Harold Garfinkelin rikkomiskokeissa, joissa keskustelun käytänteitä tarkoituksellisesti rikottiin. Erona Garfinkelin rikkomiskokeisiin on kuitenkin se, että niihin osallistuvat toivat ilmi ärtymyksensä. Tutkimuksiimme osallistuneet sen sijaan näyttävät hyväksyneen normin, jonka mukaan älypuhelimeen uppoutuminen pitää sallia, eikä toisille ole soveliasta ilmaista tästä aiheutuvaa ärtymystä. Kutsumme tätä normia älypuhelinajan kasvotyöksi.

Älypuhelinajan kasvotyö piilottaa sitä emotionaalista asemoitumista (Goodwin 2007), mitä liittyy medialaitteen tahmeuteen ja sivustakatsojan pimennon kaltaisiin vuorovaikutusilmiöihin. Samalla se osaltaan tukee harhakuvaa "multitaskaavista" ihmisistä, jotka suvereenisti toimivat kahdessa eri vuorovaikutuksessa yhtä aikaa. Eläytymismenetelmäaineistomme – sekä sarjakuvaan että naturalistiseen videoaineistoon eläytyminen – toi esiin, että puhelimen käyttö keskellä kasvokkaista keskustelua herättää tunteita. Kuitenkin tilanteessa, jossa kaikki käyttävät puhelinta, on erityisen haastavaa pyytää toisia laittamaan puhelin syrjään.

Toisinaan on esitetty, että mahdolliset medialaitteiden käytöstä juontuvat ongelmat kasvokkaisessa vuorovaikutuksessa olisivat vain välivaihe ja poistuisivat uusien sukupolvien varttuessa aikuisiksi ja heidän tottuessa erilaiseen keskustelukulttuuriin. Olemme kuitenkin jo aikaisemmassa tutkimusjulkaisussamme todenneet (Raudaskoski ym. 2017), että erityisesti vauvaikäiselle esimerkiksi sivustakatsojan pimento lienee kaikkein hankalin positio: pienelle lapselle katsekontakti häntä hoitavaan henkilöön on kehityksen kannalta erityisen tärkeää. Tuota katsekontaktia jatkuva älypuhelimen käyttö saattaa uhata. Lisäksi älypuhelimen käyttö tuo medialaitteen tahmeuden tapaisten ilmiöiden kautta epäselvyyttä sellaisiin perustavanlaatuisiin sosiaalisiin rakenteisiin, joita ihmislapsi myötäsyntyisesti arvioi ja joiden myötä hän oppii vuorovaikutuksen perussääntöjä (vrt. Levinson 2006, 44–48). Medialaitteen tahmeus ja sivustakatsojan pimento häiritsevät myös esineiden intentionaalisten tarjoumien havaitsemista. Lapselle intentionaalisten tarjoumien ymmärtäminen mahdollistaa ihmisten ja esineiden suhteiden hahmottamista sekä imitoivaa oppimista (Tomasello 1999a, 84). Kuten aiemmin tässä artikkelissa toimme esiin, sivustakatsojalle älypuhelimen intentionaaliset tarjoumat ovat kuitenkin piilossa. Sivustakatsovan pienen lapsen näkökulmasta älypuhelin näyttäytyy esineenä, jota häntä hoivaava henkilö pitää kädessään ja joka vie tuon ihmisen huomion, mutta ei paljasta käyttötarkoitustaan.

Medialaitteen tahmeuden ja sivustakatsojan pimennon sekä niihin liittyvän älypuhelinajan kasvotyön normatiivisuus ja moraalinen selontekovelvollisuus voivat vaihdella eri keskustelukonteksteissa. Perheen kesken tai työpaikalla voi olla erilaisia sääntöjä kuin vaikkapa kaveriporukassa. Jatkotutkimuksissa aiomme selvittää, onko esimerkiksi eri ikäryhmien välillä tai eri kulttuureissa, kuten eri maissa, eroa siinä, millaista vuorovaikutuksellista normistoa älypuhelinajan kasvokkaisissa keskusteluissa noudatetaan ja tuodaanko älypuhelimen käytön mahdollisesti aiheuttamaa moraalista närkästystä esiin eri tavoin näissä eri ryhmissä. Sekä asenteiden että arjen vuorovaikutuksessa tapahtuvien ilmiöiden kartoittamiseksi tulemme käyttämään aineistona sekä eri ikäryhmien että kulttuuriryhmien parissa luonnollisesti tapahtuvien vuorovaikutustilanteiden tallenteita että kyselyaineistoja, joita on kerätty eri ryhmien edustajilta.

# Kirjallisuus